\documentstyle[12pt,epsfig,graphics]{article}
\begin{document}\bibliographystyle{plain}\begin{titlepage}
\renewcommand{\thefootnote}{\fnsymbol{footnote}}\hfill\begin{tabular}{l}
HEPHY-PUB 745/01\\UWThPh-2001-37\\hep-ph/0109165\\September 2001\end{tabular}
\\[3cm]\Large\begin{center}{\bf INSTANTANEOUS BETHE--SALPETER EQUATION:
IMPROVED ANALYTICAL SOLUTION}\\\vspace{2cm}\large{\bf Wolfgang
LUCHA\footnote[1]{\normalsize\ {\em E-mail address\/}:
wolfgang.lucha@oeaw.ac.at}}\\[.3cm]\normalsize Institut f\"ur
Hochenergiephysik,\\\"Osterreichische Akademie der
Wissenschaften,\\Nikolsdorfergasse 18, A-1050 Wien, Austria\\[1cm]\large{\bf
Franz F.~SCH\"OBERL\footnote[2]{\normalsize\ {\em E-mail address\/}:
franz.schoeberl@univie.ac.at}}\\[.3cm]\normalsize Institut f\"ur Theoretische
Physik, Universit\"at Wien,\\Boltzmanngasse 5, A-1090 Wien, Austria\vfill
{\normalsize\bf Abstract}\end{center}\normalsize Studying the Bethe--Salpeter
formalism for interactions instantaneous in the rest frame of the bound
states described, we show that, for bound-state constituents of arbitrary
masses, the mass of the ground state of a given spin may be calculated almost
entirely analytically with high accuracy, without the (numerical)
diagonalization of the matrix representation obtained by expansion of the
solutions over a suitable set of basis~states.\vspace{3ex}

\renewcommand{\thefootnote}{\arabic{footnote}}\end{titlepage}

\normalsize

\section{Introduction}The Bethe--Salpeter equation in instantaneous
approximation for its interaction kernel has proven to represent a very
powerful tool for the description of bound states within the framework of
relativistic quantum field theories. In contrast to the Bethe--Salpeter
equation, the {\em instantaneous\/} Bethe--Salpeter equation (or ``Salpeter
equation'') may be formulated as an eigenvalue problem for the ``Salpeter
amplitudes'' which describe~the bound states under study; its eigenvalues are
the masses $M$ of these bound states.~In the physical sector of bound states
of {\em positive\/} norm, all mass eigenvalues $M$ are real~\cite{Lagae92I}.

In a recent series of papers
\cite{Lucha01:IBSEm=0,Lucha01:IBSEnzm,Lucha01:IBSE-C4}, we developed a
technique for the (approximate) determination of the mass eigenvalues $M$ and
the corresponding Salpeter amplitudes $\chi$ in an almost entirely analytical
way, by conversion of the instantaneous Bethe--Salpeter equation into an
eigenvalue problem for an explicitly given, analytically known matrix. Here
we demonstrate, for our (now almost ``standard'') example, how the
ground-state mass $M$ for given spin of the bound state may be obtained, with
comparable precision, even without the necessity of the numerical
diagonalization of a (rather large) matrix.

\section{$P\,C=-1$ instantaneous Bethe--Salpeter equation}In order to
facilitate an eventual comparison with the results of previous
investigations, we try to imitate the analysis of
Refs.~\cite{Lucha01:IBSEm=0,Lucha01:IBSEnzm,Lucha01:IBSE-C4} to the utmost
possible extent. First of~all, let us base~our formalism on the same two
simplifying assumptions~as~in
Refs.~\cite{Lucha01:IBSEm=0,Lucha01:IBSEnzm,Lucha01:IBSE-C4}:
\begin{enumerate}\item Every (full) fermion propagator entering in the
Bethe--Salpeter equation may~be approximated by the corresponding free one,
involving a mass parameter which then must be interpreted as some effective
mass of the bound-state constituents.\item The particles forming the bound
state under consideration have equal masses~$m.$\end{enumerate}

Moreover, let us focus our interest to fermion--antifermion bound states with
spin~$J,$ parity $P=(-1)^{J+1}$ and charge-conjugation quantum number
$C=(-1)^J$ (this means $P\,C=-1$ for all $J$), denoted by ${}^1J_J$ in the
usual spectroscopic notation. The Salpeter amplitude $\chi$ describing these
states involves two independent components, $\Psi_1$ and~$\Psi_2.$ Its
momentum-space representation $\chi({\bf k})$ is given, for two fermions of
equal masses~$m$ and internal momentum $\bf k,$ in the center-of-momentum
frame of the bound state, by$$\chi({\bf k})=\left[\Psi_1({\bf
k})\,\frac{m-\mbox{\boldmath{$\gamma$}}\cdot{\bf k}}{E(k)}+\Psi_2({\bf
k})\,\gamma^0\right]\gamma_5\ ,$$where$$E(k)\equiv\sqrt{k^2+m^2}\ ,\quad
k\equiv|{\bf k}|\ ,$$denotes the energy of a free particle of mass $m$ and
momentum~${\bf k}.$ Assume that the Salpeter amplitude $\chi$ describes bound
states with the total spin $J$ and its projection $J_3.$ Its two independent
components $\Psi_1({\bf k})$ and~$\Psi_2({\bf k})$ may then be factorized,
according~to$$\Psi_i({\bf k})=\Psi_i(k)\,{\cal Y}_{JJ_3}(\Omega)\ ,\quad
i=1,2\ ,$$into the radial wave functions $\Psi_1(k)$ and~$\Psi_2(k),$ and the
spherical harmonics ${\cal Y}_{\ell m}(\Omega)$ for angular momentum $\ell$
and its projection $m;$ the latter depend on the solid angle $\Omega,$ which
encompasses the angular variables, and satisfy the orthonormalization
condition$$\int{\rm d}\Omega\,{\cal Y}^\ast_{\ell m}(\Omega)\,{\cal
Y}_{\ell'm'}(\Omega)=\delta_{\ell\ell'}\,\delta_{mm'}\ .$$

A crucial point in the construction of any Bethe--Salpeter model for bound
states~is the determination of the Lorentz structure of the Bethe--Salpeter
kernel. According~to the analyses presented in
Refs.~\cite{Parramore95,Parramore96,Olsson95}, for an interaction potential
rising linearly~with the distance of the two bound-state constituents---as
is, for instance, frequently used~in relativistic quark models of hadrons in
order to describe the confining quark--antiquark interaction arising from
quantum chromodynamics; see, e.g., Refs.~\cite{Lucha91,Lucha92}---a kernel
with a pure time-component Lorentz-vector Dirac structure yields stable
solutions whereas a kernel with a pure Lorentz-scalar Dirac structure
certainly does not. For this reason, we have chosen in
Refs.~\cite{Lucha01:IBSEm=0,Lucha01:IBSEnzm,Lucha01:IBSE-C4} to discuss pure
time-component Lorentz-vector~kernels.

Finally, merely for notational simplicity, we confine ourselves to the case
$J=0,$~i.e., to ${}^1{\rm S}_0$ bound states, with the spin-parity-charge
conjugation assignment $J^{PC}=0^{-+}.$

For pure time-component Lorentz-vector interactions, that is, for a kernel
with the Dirac structure $\Gamma\otimes\Gamma=\gamma^0\otimes\gamma^0,$ the
instantaneous Bethe--Salpeter equation describing $P\,C=-1$
fermion--antifermion bound states with total spin $J=0$ is equivalent to~the
following set of coupled equations for the radial wave functions $\Psi_1(k)$
and~$\Psi_2(k)$ \cite{Lagae92I,Olsson95}:
\begin{eqnarray}&&2\,E(k)\,\Psi_2(k)+\int\limits_0^\infty\frac{{\rm
d}k'\,k'^2}{(2\pi)^2}\,V_0(k,k')\,\Psi_2(k')=M\,\Psi_1(k)\ ,\nonumber\\[1ex]
&&2\,E(k)\,\Psi_1(k) +\int\limits_0^\infty\frac{{\rm d}k'\,k'^2}{(2\pi)^2}\,
\frac{m^2\,V_0(k,k')+k\,k'\,V_1(k,k')}{E(k)\,E(k')}\,\Psi_1(k')=M\,\Psi_2(k)\
;\label{Eq:IBSE}\end{eqnarray}the interaction between the bound-state
constituents, described by the static potential $V(r)$ in configuration
space, enters in form of the expressions\begin{equation}V_L(k,k')\equiv
8\pi\int\limits_0^\infty{\rm d}r\,r^2\,V(r)\,j_L(k\,r)\,j_L(k'\,r)\ ,\quad
L=0,1\ ,\label{Eq:AMC-pot}\end{equation}where $j_n(z)$ ($n=0,\pm
1,\pm2,\dots$) are the spherical Bessel functions of the first
kind~\cite{Abramowitz}.

\section{Ground-state mass eigenvalue}The structure of the $P\,C=-1$
instantaneous Bethe--Salpeter equation (\ref{Eq:IBSE}) suggests~to solve this
set of equations by expressing, for bound-state mass $M\ne0,$ from the
first~of Eqs.~(\ref{Eq:IBSE}), the Salpeter component $\Psi_1(k)$ in terms of
the Salpeter component $\Psi_2(k)$,~and inserting the resulting expression
into the second of Eqs.~(\ref{Eq:IBSE}). By this procedure the set of
equations (\ref{Eq:IBSE}) is reduced to an eigenvalue equation for
$\Psi_2(k)$ with $M^2$ as eigenvalue:
\begin{eqnarray}M^2\,\Psi_2(k)&=&4\,E^2(k)\,\Psi_2(k)
+2\,E(k)\int\limits_0^\infty\frac{{\rm d}k'\,k'^2}{(2\pi)^2}\,
V_0(k,k')\,\Psi_2(k')\nonumber\\[1ex]&+&2\int\limits_0^\infty\frac{{\rm
d}k'\,k'^2}{(2\pi)^2}\,\frac{m^2\,V_0(k,k')+k\,k'\,V_1(k,k')}{E(k)}\,\Psi_2(k')
\label{Eq:IBSE-M2}\\[1ex]&+&\int\limits_0^\infty\frac{{\rm
d}k'\,k'^2}{(2\pi)^2}\,\frac{m^2\,V_0(k,k')+k\,k'\,V_1(k,k')}{E(k)\,E(k')}\,
\int\limits_0^\infty\frac{{\rm d}k''\,k''^2}{(2\pi)^2}\,V_0(k',k'')\,
\Psi_2(k'')\ .\nonumber\end{eqnarray}

Our goal is to find an as far as possible analytic albeit approximate
characterization of the bound-state masses $M.$ To this end, we rely on an
approximate description~of~the analyzed bound states by (trial) states
$|\phi\rangle$ which involve a real variational parameter~$\mu.$

\newpage

Our choice for $|\phi\rangle$ makes use of the exponential; our
$|\phi\rangle$ is defined in terms of~its~real configuration-space or
momentum-space representation $\phi(r)$ or $\phi(p),$ respectively,
by$$\phi(r)=2\,\mu^{3/2}\exp(-\mu\,r)\
,\quad\phi(p)=\sqrt{\frac{2}{\pi}}\,\frac{4\,\mu^{5/2}}{(p^2+\mu^2)^2}\
,$$Normalizability of the Hilbert-space states $|\phi\rangle$ requires $\mu$
to be strictly positive: $\mu>0.$ These trial functions $\phi(r)$ and
$\phi(p)$ are related by the Fourier--Bessel transformations
\begin{equation}\phi(r)=\sqrt{\frac{2}{\pi}}\int\limits_0^\infty{\rm
d}p\,p^2\,j_0(p\,r)\,\phi(p)\
,\quad\phi(p)=\sqrt{\frac{2}{\pi}}\int\limits_0^\infty{\rm
d}r\,r^2\,j_0(p\,r)\,\phi(r)\ ,\label{Eq:FBT}\end{equation}and they satisfy
the normalization conditions$$\int\limits_0^\infty{\rm
d}r\,r^2\,\phi^2(r)=\int\limits_0^\infty{\rm d}p\,p^2\,\phi^2(p)=1\ .$$The
latter feature allows us to extract, from Eq.~(\ref{Eq:IBSE-M2}), the
eigenvalue $M^2$ in closed form:
\begin{eqnarray}M^2&=&4\int\limits_0^\infty{\rm
d}k\,k^2\,E^2(k)\,\phi^2(k)+\frac{2}{(2\pi)^2}
\int\limits_0^\infty{\rm d}k\,k^2\,E(k)\,\phi(k)
\int\limits_0^\infty{\rm d}k'\,k'^2\,V_0(k,k')\,\phi(k')\nonumber\\[1ex]
&+&\frac{2\,m^2}{(2\pi)^2}
\int\limits_0^\infty\frac{{\rm d}k\,k^2}{E(k)}\,\phi(k)
\int\limits_0^\infty{\rm d}k'\,k'^2\,V_0(k,k')\,\phi(k')\nonumber\\[1ex]
&+&\frac{2}{(2\pi)^2}
\int\limits_0^\infty\frac{{\rm d}k\,k^3}{E(k)}\,\phi(k)
\int\limits_0^\infty{\rm d}k'\,k'^3\,V_1(k,k')\,\phi(k')\nonumber\\[1ex]
&+&\frac{m^2}{(2\pi)^4}
\int\limits_0^\infty\frac{{\rm d}k\,k^2}{E(k)}\,\phi(k)
\int\limits_0^\infty\frac{{\rm d}k'\,k'^2}{E(k')}\,V_0(k,k')
\int\limits_0^\infty{\rm d}k''\,k''^2\,V_0(k',k'')\,\phi(k'')\nonumber\\[1ex]
&+&\frac{1}{(2\pi)^4}
\int\limits_0^\infty\frac{{\rm d}k\,k^3}{E(k)}\,\phi(k)
\int\limits_0^\infty\frac{{\rm d}k'\,k'^3}{E(k')}\,V_1(k,k')
\int\limits_0^\infty{\rm d}k''\,k''^2\,V_0(k',k'')\,\phi(k'')\
.\label{Eq:M2}\end{eqnarray}

Now, the angular-momentum component $V_1(k,k')$ of the interaction potential
$V(r)$ involves---according to its definition (\ref{Eq:AMC-pot})---the
spherical Bessel function $j_1(z)$ of order~1. Consequently, in order to deal
with those terms in Eq.~(\ref{Eq:M2}) which involve the component
$V_1(k,k'),$ it is advisable to introduce, in addition, states $|\psi\rangle$
that are related to angular momentum 1; the configuration-space and
momentum-space representations of $|\psi\rangle$~are then related by a
Fourier--Bessel transformation similar to Eq.~(\ref{Eq:FBT}) but
involving~$j_1(z)$:$$\psi(r)=\frac{2\,\mu^{5/2}}{\sqrt{3}}\,r\exp(-\mu\,r)\
,\quad\psi(p)=-{\rm i}\,\sqrt{\frac{2}{3\,\pi}}\,\frac{16\,\mu^{7/2}\,p}
{(p^2+\mu^2)^3}\ ,$$with$$\psi(r)={\rm
i}\,\sqrt{\frac{2}{\pi}}\int\limits_0^\infty{\rm
d}p\,p^2\,j_1(p\,r)\,\psi(p)\ ,\quad\psi(p)=-{\rm
i}\,\sqrt{\frac{2}{\pi}}\int\limits_0^\infty{\rm
d}r\,r^2\,j_1(p\,r)\,\psi(r)\ .$$These further functions $\psi(r)$ and
$\psi(p)$ are also normalized to unity, that is,
they~satisfy$$\int\limits_0^\infty{\rm
d}r\,r^2\,\psi^2(r)=\int\limits_0^\infty{\rm d}p\,p^2\,|\psi(p)|^2=1\ .$$

\newpage

The evaluation of the kinetic energy in the expression (\ref{Eq:M2}) for
$M^2$ is straightforward:$$\int\limits_0^\infty{\rm
d}k\,k^2\,E^2(k)\,\phi^2(k)=\int\limits_0^\infty{\rm
d}k\,k^4\,\phi^2(k)+m^2=\mu^2+m^2\ .$$However, in order to evaluate the
interaction terms in Eq.~(\ref{Eq:M2}), we have to approximate the various
expressions entering in the integrands of these terms in the following
way:\begin{eqnarray*}E(k)\,\phi(k)&=&b\,\phi(k)\ ,\\[1ex]
\frac{k}{E(k)}\,\phi(k)&=&c\,\psi(k)\ ,\\[1ex]k\,\phi(k)&=&d\,\psi(k)\
,\\[1ex]\frac{1}{E(k)}\,\phi(k)&=&e\,\phi(k)\ .\end{eqnarray*}Expressed in
terms of the coefficients $b,$ $c,$ $d,$ and $e$ defined above, the square
(\ref{Eq:M2}) of~the bound-state mass eigenvalue $M$ of the instantaneous
Bethe--Salpeter equation~(\ref{Eq:IBSE})~may then be traced back to the
expectation
values\begin{eqnarray*}V^{(0)}&\equiv&\langle\phi|V(r)|\phi\rangle=
\int\limits_0^\infty{\rm d}r\,r^2\,V(r)\,\phi^2(r)\ ,\\[1ex]
V^{(1)}&\equiv&\langle\psi|V(r)|\psi\rangle=
\int\limits_0^\infty{\rm d}r\,r^2\,V(r)\,\psi^2(r)\end{eqnarray*}of the
interaction potential $V(r)$ with respect to the approximation functions
$\phi$ and~$\psi$:
\begin{eqnarray}M^2&=&4\left(m^2+\mu^2\right)+2\left(b+m^2\,e\right)V^{(0)}
+2\,c^\ast\,d\,V^{(1)}\nonumber\\[1ex]&+&m^2\,e^2\left(V^{(0)}\right)^2
+|c|^2\,V^{(0)}\,V^{(1)}\ .\label{Eq:M2-1x1}\end{eqnarray}For the factors
$b,$ $c,$ $d,$ and $e$ analytic expressions may be found; the latter will
obviously involve the mass $m$ of the bound-state constituents and the
variational parameter~$\mu$:
\begin{eqnarray*}b&=&\int\limits_0^\infty{\rm d}k\,k^2\,E(k)\,\phi^2(k)
=\mu\,x^4\left[\frac{x^2-2}{(x^2-1)^{5/2}}+\frac{2^6}{15\,\pi\,x^8}\,
F\left(2,4;\frac{7}{2};\frac{1}{x^2}\right)\right],\\[1ex]
c&=&\int\limits_0^\infty\frac{{\rm d}k\,k^3}{E(k)}\,\psi^\ast(k)\,\phi(k)
=\frac{{\rm i}\,x^4}{2\,\sqrt{3}}\left[\frac{3\,x^4-16\,x^2+48}
{(x^2-1)^{9/2}}-\frac{2^{11}}{15\,\pi\,x^{10}}\,
F\left(3,5;\frac{7}{2};\frac{1}{x^2}\right)\right],\\[1ex]
d&=&\int\limits_0^\infty{\rm d}k\,k^3\,\psi^\ast(k)\,\phi(k) ={\rm
i}\,\frac{\sqrt{3}}{2}\,\mu\ ,\\[1ex] e&=&\int\limits_0^\infty\frac{{\rm
d}k\,k^2}{E(k)}\,\phi^2(k)
=\frac{x^2}{\mu}\left[\frac{x^4-4\,x^2+8}{(x^2-1)^{7/2}}-
\frac{2^8}{15\,\pi\,x^8}\,F\left(3,4;\frac{7}{2};\frac{1}{x^2}\right)\right]
,\end{eqnarray*}with the Gauss hypergeometric series $F$, defined, in terms
of the gamma function~$\Gamma,$~by \cite{Abramowitz}
$$F(u,v;w;z)=\frac{\Gamma(w)}{\Gamma(u)\,\Gamma(v)}\,\sum_{n=0}^\infty\,
\frac{\Gamma(u+n)\,\Gamma(v+n)}{\Gamma(w+n)}\,\frac{z^n}{n!}\ ,$$and the
abbreviation$$x\equiv\frac{m}{\mu}$$for the ratio of the mass parameters
involved. It is rather easy to convince oneself that these expressions for
the coefficients $b,$ $c,$ $d,$ and $e$ reduce, in the (ultrarelativistic)
limit $m\to0,$ to the lowest entries of the corresponding matrices given in
Ref.~\cite{Lucha01:IBSEm=0}, where~the instantaneous Bethe--Salpeter equation
has been studied for the simpler special case~of vanishing masses of all
bound-state constituents, and, for the particular value $\mu=m,$ to the
lowest entries of the corresponding (similarly defined) matrices given in
Ref.~\cite{Lucha01:IBSEnzm}, where, because of the nonvanishing masses of the
bound-state constituents, the desired analytical treatment forced us to fix
the value of the variational parameter $\mu$ to $\mu=m.$

The necessarily numerical optimization of the right-hand side of the analytic
result (\ref{Eq:M2-1x1}) with respect to the variational parameter $\mu$ then
yields a first approximation to~the masses $M$ of bound states described by
the instantaneous Bethe--Salpeter equation~(\ref{Eq:IBSE}).

\section{Accuracy of the approximation}The decisive question clearly is
whether it is possible to achieve a satisfactory accuracy of the bound-state
mass $M$ calculated from the approximation represented by
Eq.~(\ref{Eq:M2-1x1}). The answer to this question will, of course, depend on
the interaction potential $V(r).$

The general expressions for the expectation values of $V(r)$ for arbitrary
power-law potentials may be deduced from
Refs.~\cite{Lucha97,Lucha98O,Lucha98D}. Here we study as a simple but,~from
the physical point of view, nevertheless relevant example the case of a
linear potential:$$V(r)=\lambda\,r\ ,\quad\lambda>0\ .$$Inserting the
required expectation values $V^{(0)}$ and $V^{(1)}$ of this potential (which
may~be found, e.g., in Refs.~\cite{Lucha01:IBSEm=0,Lucha01:IBSEnzm}),
$$V^{(0)}=\frac{3\,\lambda}{2\,\mu}\ ,\quad
V^{(1)}=\frac{5\,\lambda}{2\,\mu}\ ,$$into our general formula
(\ref{Eq:M2-1x1}) for the mass squared of the lowest ${}^1{\rm S}_0$ bound
state yields\begin{equation}
M^2=4\left(m^2+\mu^2\right)+\frac{3\,\lambda}{\mu}\left(b+m^2\,e\right)
+\frac{5\,\lambda}{\mu}\,c^\ast\,d+\frac{3\,\lambda^2}
{4\,\mu^2}\left(3\,m^2\,e^2+5\,|c|^2\right).\label{Eq:M2-LP}\end{equation}
Minimization of this expression with respect to $\mu,$ for
$m=0.1\;\mbox{GeV}$ and $\lambda=0.2\;\mbox{GeV}^2,$ gives
$M=1.703\;\mbox{GeV};$ this is only $2.5\%$ larger than the ``exact'' value
$M=1.661\;\mbox{GeV}$ computed in Ref.~\cite{Lucha01:IBSEnzm} by the
diagonalization of a $25\times 25$ matrix (cf.\ Table~1 of
Ref.~\cite{Lucha01:IBSEnzm}). Table~\ref{Tab:Comparison} compares the squared
masses predicted by Eq.~(\ref{Eq:M2-LP}) with the findings presented in
Table~1 of Ref.~\cite{Parramore95}. Again, the relative errors are only of
the order of a few percent.

\begin{table}[ht]\caption{Lowest mass eigenvalues $M$ of the instantaneous
Bethe--Salpeter equation, with a time-component Lorentz-vector interaction
kernel, describing bound states~with spin-parity-charge conjugation
assignment $J^{PC}=0^{-+}$ of two spin-$\frac{1}{2}$ fermions of mass $m,$
which experience a confining interaction described by a linear potential
$V(r)=\lambda\,r$ with slope $\lambda=0.29\;\mbox{GeV}^2.$ The values of
$M^2$ arising, for various values of the mass $m$ of the bound-state
constituents, within the present approach (third column) differ by~a few
percent (fourth column) from the values of $M^2$ obtained in
Ref.~\cite{Parramore95} by expanding the Salpeter amplitude $\chi$ over a
basis of 25 harmonic-oscillator eigenfunctions (second column). Moreover, the
relative errors decrease with increasing constituents' mass
$m.$}\label{Tab:Comparison}
\begin{center}\begin{tabular}{ccccc}\hline\hline\\[-1.5ex]
\multicolumn{1}{c}{$m$ [MeV]}&\multicolumn{1}{c}{$M^2$ [GeV${}^2$]}&
\multicolumn{1}{c}{$M^2$ [GeV${}^2$]}&\multicolumn{1}{c}{Relative Error}&
\multicolumn{1}{c}{Relative Error}\\
&\multicolumn{1}{c}{(Ref.~\cite{Parramore95})}&\multicolumn{1}{c}{(present
work)}&\multicolumn{1}{c}{of $M^2$ [\%]}&\multicolumn{1}{c}{of $M$ [\%]}
\\[1ex]\hline\\[-1.5ex]
300&4.357&4.568&4.8&2.4\\500&5.255&5.478&4.2&2.1\\900&8.248&8.515&3.2&1.6
\\[1ex]\hline\hline\end{tabular}\end{center}\end{table}

\section{Summary and conclusion}The demand to derive, also for the case of
massive bound-state constituents, analytical matrix representations of the
instantaneous Bethe--Salpeter equation led us, in
Ref.~\cite{Lucha01:IBSEnzm}, to identify the variational parameter in the
trial states used in Ref.~\cite{Lucha01:IBSEm=0} with the mass~$m$ of the
bound-state constituents. However, the rate of convergence of the
corresponding bound-state masses $M$ with increasing matrix size is, because
of this identification,~for {\em small\/} but nonvanishing masses $m$ of the
bound-state constituents not extremely rapid. For a first idea of the
location of $M,$ one might want to avoid matrix diagonalizations.

\newpage\noindent Here, by applying the same variational technique as used in
Ref.~\cite{Lucha01:IBSEm=0}, we have been~able to demonstrate that, at least
for ground states, the bound-state mass $M$ may be found with reasonable
accuracy, certainly sufficient for the initial steps of a fitting procedure.

\end{document}